\documentclass{vldb20_arxiv_vldb}

\pdfoutput=1 

\usepackage{graphicx}
\usepackage{balance}  

\vldbTitle{LMFAO: An Engine for Batches of Group-By Aggregates}
\vldbAuthors{Maximilian Schleich and Dan Olteanu}
\vldbDOI{https://doi.org/110.14778/3415478.3415515}
\vldbVolume{13}
\vldbNumber{12}
\vldbYear{2020}

\usepackage{color, xcolor}
\colorlet{dred}{red!80!black}
\colorlet{dgreen}{green!50!black}
\colorlet{dorange}{orange!90!black}

\definecolor{lightred}{rgb}{1,0.7,0.7}
\definecolor{lightgreen}{rgb}{0.7,1,0.7}
\definecolor{lightblue}{rgb}{0.7,0.7,1}
\definecolor{lightblue}{rgb}{0.55,0.72,0.97}
\definecolor{lightyellow}{rgb}{1,1,0.4} 
\definecolor{lightorange}{rgb}{1,0.7,0.5}

\definecolor{light}{gray}{0.85}
\definecolor{heavy}{gray}{0.35}
\definecolor{goodgreen}{rgb}{0.1, 0.5, 0.1}

\definecolor{airforceblue}{rgb}{0.36, 0.54, 0.66}
\definecolor{amethyst}{rgb}{0.6, 0.4, 0.8}
\definecolor{amaranth}{rgb}{0.9, 0.17, 0.31}
\definecolor{amber}{rgb}{1.0, 0.49, 0.0}
\definecolor{applegreen}{rgb}{0.55, 0.71, 0.0}
\definecolor{oxfordblue}{rgb}{0, 0.33, 0.71}
\definecolor{forestgreen}{rgb}{0.13, 0.55, 0.13}
\definecolor{auburn}{rgb}{0.43, 0.21, 0.1}

\usepackage{tikz,tikz-qtree,tikz-qtree-compat}
\usetikzlibrary{fit,topaths,calc,shapes,decorations,chains,scopes,arrows,calc,positioning,matrix,backgrounds}

\usepackage{subcaption}
\usepackage{url}

\usepackage{bm}
\usepackage[export]{adjustbox}

\tikzstyle{path} = [->,double,rounded corners=.1cm]
\tikzstyle{data} = [draw, rectangle, rounded corners = .07cm, align=center, inner sep = .2cm, outer sep = .1 cm]
\tikzstyle{processor} = [draw, ellipse, align=center, inner sep = .1cm, outer sep = .1 cm]
\colorlet{clr_outofDB}{red!80!black}
\colorlet{clr_inDB}{green!50!black}
\colorlet{clr_inDB_FD}{blue}

\newcommand{\drawtable}[5]
{
   \begin{scope}[shift={(-#1/2,-#2/2)}]
      \pgfmathsetmacro{\cellwidth}{#1/#3}
      \pgfmathsetmacro{\cellheight}{#2/#4}
      \draw[fill=#5!30] (0, #2) rectangle (#1, #2-\cellheight);
      \foreach \i in {1,...,#3}
         \draw[thin, #5!70] (\cellwidth*\i, 0) -- (\cellwidth*\i, #2);
      \foreach \i in {1,...,#4}
         \draw[thin, #5!70] (0, \cellheight*\i) -- (#1, \cellheight*\i);
      \draw[#5] (0, #2-\cellheight) -- (#1, #2-\cellheight);
      \draw[thick, #5] (0, 0) rectangle (#1, #2);
   \end{scope}
}

\newcommand{\drawcylinder}[4]
{
   \begin{scope}[shift={(0,#3/2)}]
      \draw[#4] (0, #2/2) ellipse (#1/2 and #3);
      \draw[#4] (#1/2, -#2/2) arc (0:-180:#1/2 and #3);
      \draw[#4] (-#1/2, #2/2) -- (-#1/2, -#2/2);
      \draw[#4] (#1/2, #2/2) -- (#1/2, -#2/2);
   \end{scope}
}

\newcommand{\drawfilter}[2]
{
   \begin{scope}[scale=#1/7, shift={(-3.5,3)}]
      \draw[#2] (0,0)--(7, 0)--(4, -3)--(4,-5)--(3,-6)--(3,-3)--cycle;
   \end{scope}
}

\newcommand{\pluseq}{\mathrel{+}=}
\newcommand{\norm}[1]{\left\|#1\right\|}

\newcommand{\gv}[1]{\ensuremath{\mbox{\boldmath$ #1 $}}} 
\newcommand{\grad}[1]{\gv{\nabla} #1} 

\newcommand{\inner}[1]{\left\langle #1 \right\rangle}

\renewcommand{\vec}[1]{\ensuremath\boldsymbol{#1}}

\begin{document}


\title{LMFAO: An Engine for Batches of Group-By Aggregates\\[0.5em] {\Large\bf Layered Multiple Functional Aggregate Optimization}}



%
%
%
%

\numberofauthors{2} 

\author{
%
%
\alignauthor Maximilian Schleich\\[0.5ex]
       \affaddr{University of Washington\\}
       \email{schleich@cs.washington.edu}
\alignauthor Dan Olteanu\\[0.5ex]
       \affaddr{University of Zurich\\}
       \email{olteanu@ifi.uzh.ch}
}

\date{}

\maketitle

\begin{abstract}

  \vspace*{1em}

  LMFAO is an in-memory optimization and execution engine for large batches of
  group-by aggregates over joins. Such database workloads capture
  the data-intensive computation of a variety of data science applications. 

  We demonstrate LMFAO for three popular models: ridge linear
  regression with batch gradient descent, decision trees with CART, and clustering with Rk-means.
\end{abstract}

\vspace*{1em}

\section{LMFAO's Approach to Learning over Relational Databases}
\label{sec:intro}

\vspace*{1em}


LMFAO is born out of the necessity to efficiently support ubiquitous data
science workloads that involve learning models over relational
queries~\cite{lmfao}.  From a database perspective, the data-intensive
computation required by such workloads can be expressed as batches of group-by
aggregates over the join of the underlying database relations. By tightly
integrating the query processing and the learning tasks, LMFAO can outperform
mainstream solutions based on TensorFlow and scikit-learn over Pandas by several
orders of magnitude~\cite{lmfao}. Such workloads pose new challenges to
relational data processing engines as they require the computation of hundreds
to thousands of similar yet distinct group-by aggregates over the 
natural join of database relations. Prior experiments with commercial and
open-source database systems including MonetDB and PostgreSQL confirm that these
challenges are not yet addressed satisfactorily by existing database
technology~\cite{lmfao}.
  
To address these challenges, LMFAO puts forward a layered architecture of optimizations that chiefly target computation sharing at all data processing stages, factoring out repeated computation, and code specialization.

We demonstrate LMFAO for three popular models: ridge linear regression using batch gradient descent, decision trees using CART~\cite{cart84}, and clustering using Rk-means~\cite{Rkmeans:AISTATS:2020}. We learn them over commercial (Retailer~\cite{lmfao}; 84M tuples) and public (Favorita~\cite{favorita}; 120M tuples) multi-relational datasets, which have been previously used for benchmarking LMFAO and its predecessors~\cite{SOC:SIGMOD:16,ANNOS:DEEM:18,lmfao,Rkmeans:AISTATS:2020}.  Prior and on-going work by the authors (\url{https://fdbresearch.github.io}) showed that the LMFAO approach is useful for a variety of further discriminative and generative models, e.g., Generalized Linear Models, Support Vector Machines, (robust) PCA, Factorization Machines, and Sum-Product Networks.

\vfill


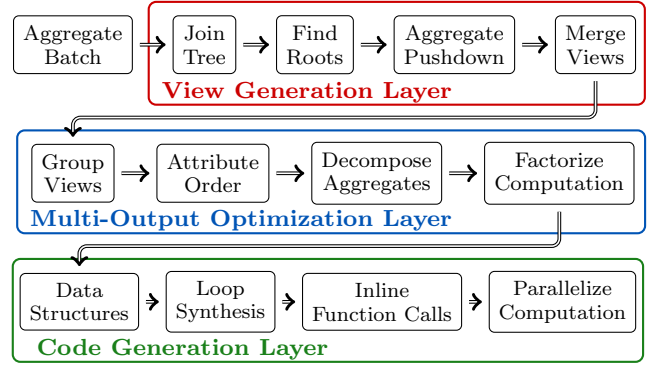
\begin{figure}[t]
  \centering \small 
  \begin{tikzpicture}[scale=0.95, xscale=0.6, yscale=0.6]
    \tikzstyle{data} = [draw, rectangle, rounded corners = .07cm, align=center,
    inner sep = .13cm, outer sep = .1 cm]
    \tikzstyle{path} = [->, double, rounded corners=.1cm]

    \node[data,scale=1] (aggs) {Aggregate\\ Batch};

    \node[data,scale=1,anchor=west] at ($(aggs.east) + (0.6,0)$) (td) {Join\\ Tree};
    \node[data,scale=1,anchor=west] at ($(td.east) + (0.6,0)$) (root) {Find \\ Roots};
    \node[data,scale=1,anchor=west] at ($(root.east) + (0.6,0)$) (pushdown) {Aggregate\\ Pushdown};
    \node[data,scale=1,anchor=west] at ($(pushdown.east) + (0.6,0)$) (mergeview) {Merge\\ Views};

    \draw[draw=dred, rounded corners = .1cm, thick]
    ($(td)+(-1.3,1)$) rectangle ($(mergeview)+(1.15,-1.4)$);
    \node[color=dred, anchor=west, scale=1.1] at ($(td)+(-1.2,-1.1)$)
    {\bf View Generation Layer};

    \node[data,scale=1] at ($(aggs) + (0,-3)$) (group) {Group\\ Views};
    \node[data,scale=1,anchor=west] at ($(group.east) + (0.65,0)$) (order) {Attribute\\ Order};
    \node[data,scale=1,anchor=west] at ($(order.east) + (0.65,0)$) (viewreg) {Decompose\\ Aggregates};
    \node[data,scale=1,anchor=west] at ($(viewreg.east) + (0.65,0)$) (aggreg) {Factorize\\ Computation};

    \draw[draw=oxfordblue, rounded corners = .1cm, thick]
    ($(group)+(-1.3,1)$) rectangle ($(aggreg)+(2,-1.4)$);
    \node[color=oxfordblue, anchor=west, scale=1.1] at ($(group)+(-1.2,-1.1)$)
    {\bf Multi-Output Optimization Layer};

    \node[data,scale=.98] at ($(aggs) + (.15,-6)$) (ds) {Data\\ Structures};
    \node[data,scale=.98,anchor=west] at ($(ds.east) + (0.3,0)$) (loop) {Loop\\ Synthesis};
    \node[data,scale=.98,anchor=west] at ($(loop.east) + (0.3,0)$) (parallel) {Inline \\ Function Calls};
    \node[data,scale=.98,anchor=west] at ($(parallel.east) + (0.3,0)$) (comp) {Parallelize\\ Computation};

    \draw[draw=goodgreen, rounded corners = .1cm, thick]
    ($(ds.west)+(0,1)$) rectangle ($(comp.east)+(0.0,-1.4)$);
    \node[color=goodgreen, anchor=west, scale=1.1] at ($(ds)+(-1.2,-1.1)$)
    {\bf Code Generation Layer};

    \draw[path] (aggs) -- (td);
    \draw[path] (td) -- (root);
    \draw[path] (root) -- (pushdown);
    \draw[path] (pushdown) -- (mergeview);

    \draw[path] (mergeview) -- ++(0,-1.7) -| (group);

    \draw[path] (group) -- (order);
    \draw[path] (order) -- (viewreg);
    \draw[path] (viewreg) -- (aggreg);

    \draw[path] (aggreg) -- ++(0,-1.7) -| (ds);

    \draw[path] (ds) -- (loop);
    \draw[path] (loop) -- (parallel);
    \draw[path] (parallel) -- (comp);
  \end{tikzpicture}

  \caption{The layers of LMFAO.}
  \label{fig:lmfao-layers}
  \vspace*{1em}
\end{figure}

\section{LMFAO by Example}
\label{sec:lmfao}

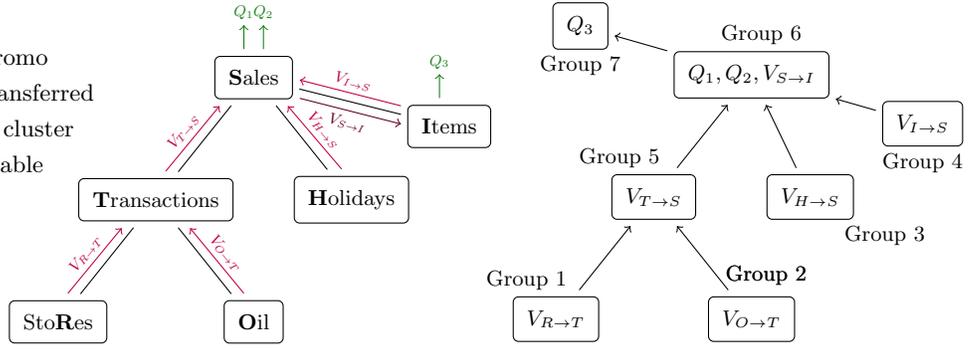
\begin{figure*}[t] 
  \begin{subfigure}[t]{0.35\textwidth}
    \begin{minipage}{0.35\textwidth}
      \begin{align*}
        &\text{{\bf Sales:} \underline{date}, \underline{store}, \underline{item}, units, promo}\\
        &\text{{\bf Holidays:} \underline{date}, htype, locale, transferred}\\
        &\text{{\bf StoRes:} \underline{store}, city, state, stype, cluster}\\
        &\text{{\bf Items:} \underline{item}, family, class, perishable}\\
        &\text{{\bf Transactions:} \underline{date}, \underline{store}, txns}\\
        &\text{{\bf Oil:} \underline{date}, price}\\
        & \\
      \end{align*}
    \end{minipage}
  \end{subfigure}
  \begin{subfigure}[t]{0.36\textwidth}
  \begin{minipage}{6cm}\hspace*{-1.5cm}
      \begin{tikzpicture}[yscale = 0.65, xscale=0.65, every node/.style={transform shape}]
        \node[data,scale=1.4] (sales) {\textbf{S}ales};
        \node[data,scale=1.4] at($(sales)+(-2,-2.5)$) (trans) {\textbf{T}ransactions};
        \node[data,scale=1.4] at($(trans)+(-2,-2.5)$) (store) {Sto\textbf{R}es};
        \node[data,scale=1.4] at($(trans)+(2,-2.5)$) (oil) {\textbf{O}il};        
        \node[data,scale=1.4] at($(sales)+(4,-1)$) (item) {\textbf{I}tems};
        \node[data,scale=1.4] at($(sales)+(2,-2.5)$) (holi) {\textbf{H}olidays};

        \draw (sales) -- (trans); 
        \draw (trans) -- (store); 
        \draw (trans) -- (oil);
        \draw (sales) -- (item);
        \draw (sales) -- (holi);

        \draw[->,color=purple] ($(trans.north)+(0.2, 0.01)$) -- ($(sales.south)+(-0.7, -0.01)$)
        node[midway,above,sloped,scale=1]{$V_{T\rightarrow S}$};
        
        \draw[->,color=purple] ($(store.north)+(0.2, 0.01)$) -- ($(trans.south)+(-0.7, -0.01)$) node[midway,above,sloped,scale=1]{{\color{purple}$V_{R\rightarrow T}$}};xs

        \draw[->,color=purple] ($(oil.north)+(-0.2, 0.01)$) -- ($(trans.south)+(0.7, -0.01)$)
        node[midway,above,sloped,scale=1]{$V_{O\rightarrow T}$};

        \draw[->,color=purple] ($(holi.north)+(-0.2, 0.01)$) -- ($(sales.south)+(0.7, -0.01)$)
        node[midway,above,sloped,scale=1]{$V_{H\rightarrow S}$};

        \draw[->,color=purple] ($(item.north west)+(0, -0.15)$) -- ($(sales.east)+(0.01, -0.05)$) node[midway,above,sloped,scale=1]{{\color{purple}$V_{I\rightarrow S}$}}; 
        
        \draw[->,color=purple!50!black] ($(sales.south east)+(0.01, 0.15)$) -- ($(item.west)+(0, 0.05)$)
        node[midway,below,sloped,scale=1]{$V_{S\rightarrow I}$};

        \draw[->,color=goodgreen] ($(sales.north)+(-0.2, 0.01)$) -- ($(sales.north)+(-0.2, 0.5)$)
        node[anchor=south]{$Q_1$};
        \draw[->,color=goodgreen] ($(sales.north)+(0.2, 0.01)$) -- ($(sales.north)+(0.2, 0.5)$)
        node[anchor=south]{$Q_2$};

        \draw[->,color=goodgreen] ($(item.north)+(-0.2, 0.01)$) -- ($(item.north)+(-0.2, 0.5)$)
        node[anchor=south]{$Q_3$};

      \end{tikzpicture}
    \end{minipage}
  \end{subfigure}
  \begin{subfigure}[t]{0.3\textwidth}
    \begin{minipage}{5cm}\hspace*{-1.7cm}
      \begin{tikzpicture}[yscale = 0.65, xscale=0.65, every node/.style={transform shape}]
        \node[data,scale=1.4] (sales) {$Q_1, Q_2, V_{S\rightarrow I}$};
        \node[data,scale=1.4] at($(sales)+(-2,-2.5)$) (trans) {$V_{T\rightarrow S}$};
        \node[data,scale=1.4] at($(trans)+(-2,-2.5)$) (store) {$V_{R\rightarrow T}$};
        \node[data,scale=1.4] at($(trans)+(2,-2.5)$) (oil) {$V_{O\rightarrow T}$};        
        \node[data,scale=1.4] at($(sales)+(3.5,-1)$) (item) {$V_{I\rightarrow S}$};
        \node[data,scale=1.4] at($(sales)+(1.2,-2.5)$) (holi) {$V_{H\rightarrow S}$};
        \node[data,scale=1.4] at($(sales)+(-3.5,1)$) (g7) {$Q_3$};

        \draw[<-] (sales) -- (trans); 
        \draw[<-] (trans) -- (store); 
        \draw[<-] (trans) -- (oil); 
        \draw[<-] (sales) -- (item);
        \draw[<-] (sales) -- (holi);
        \draw[->] (sales) -- (g7);

        \node[anchor=south] at ($(sales.north)+(0.2, 0.5)$) {};

        \node[scale=1.4]  at ($(sales.north)+(0.2,0.2)$) (a) {Group 6};
        \node[scale=1.4]  at ($(trans.north)+(-0.7,0.2)$) (a) {Group 5};
        \node[scale=1.4]  at ($(store.north)+(-0.6,0.2)$) (a) {Group 1};
        \node[scale=1.4]  at ($(oil.north)+(0.3,0.3)$) (a) {Group 2};
        \node[scale=1.4]  at ($(item.south)+(0,-0.2)$) (a) {Group 4};
        \node[scale=1.4]  at ($(oil.north)+(0.3,0.3)$) (a) {Group 2};
        \node[scale=1.4]  at ($(holi.south east)+(0.5,-0.2)$) (a) {Group 3};
        \node[scale=1.4]  at ($(g7.south)+(0,-0.2)$) (a) {Group 7};

      \end{tikzpicture}
    \end{minipage}
  \end{subfigure}
  \caption{(left) The schema for the Favorita dataset. (middle) A join tree for
    this schema with directional views and three queries, partitioned in 7
    groups. (right) The dependency graph of the groups of views and output queries.}
  \label{fig:join-tree-favorita}
\end{figure*}

Figure~\ref{fig:lmfao-layers} depicts the layered architecture of LMFAO.
We next explain these layers using an example with three group-by aggregate queries over the Favorita dataset~\cite{favorita}, whose schema is depicted in Figure~\ref{fig:join-tree-favorita}; $\mathtt{D}$ is the natural join $\mathtt{S}\Join \mathtt{T} \Join \mathtt{R}\Join \mathtt{O}\Join \mathtt{H}\Join \mathtt{I}$ of all relations, $h$ and $g$ are user-defined aggregate functions returning numerical values.

\vspace*{-.5em}
{\small
  \begin{align*}
    &\mathtt{Q_1}\texttt{\,=\,SELECT\,SUM(units)\,FROM\,D} \\
    &\mathtt{Q_2}\texttt{\,=\,SELECT\,store,SUM(g(item)*h(date))\,FROM\,D\,GROUP\,BY\,store}\\
    &\mathtt{Q_3}\texttt{\,=\,SELECT\,class,SUM(units*price)\,FROM\,D\,GROUP\,BY\,class}
  \end{align*}
}
\vspace*{-.5em}

The {\bf\color{dred} View Generation} layer takes the batch of
 queries, the database schema, and cardinality constraints (e.g., sizes of relations and attribute domains) and produces one query plan for all queries. The backbone of this plan is a join tree. 
 
In the absence of group-by clauses, LMFAO computes each query $Q$ in one bottom-up pass over the join tree by decomposing it into views computed along each edge in the tree. The view at an edge going out of a node $n$ computes the subquery that is the restriction of $Q$ to the attributes in the subtree rooted at $n$ and over the join of the relation at $n$ and of the views at the incoming edges of $n$. 

In the presence of group-by clauses, these views would have to carry the values for the group-by attributes along the paths from leaves to the root of the join tree. These views may be large and require significant compute time. 

To alleviate this problem, one approach is to use a different join tree for each query so as to minimize the sizes of these views, e.g., by choosing a tree whose root has the group-by attributes of the query with the largest domains. This approach is however expensive as it would require to recompute the joins for each query. There is also no sharing of computation across the queries.
 
Instead, LMFAO compromises between the two aforementioned approaches. It uses one join tree for all queries, but assigns one root per query (using a simple heuristic~\cite{lmfao}). Each query is thus decomposed into one view per edge in the join tree in a top-down traversal starting at its assigned root. This means that some edges may be traversed in both directions. This can reduce the sizes of the views and increase the sharing of their computation, thereby reducing the overall compute time. In our example, we choose Sales as root for $\mathtt{Q_1}$ and $\mathtt{Q_2}$, and Items as root for $\mathtt{Q_3}$.

After each query is decomposed into views at edges in the join tree, LMFAO merges views whenever they have the same direction and group-by attributes. A single view may thus be used for several queries. Figure~\ref{fig:join-tree-favorita} (middle) depicts the merged views for $\mathtt{Q_1},\mathtt{Q_2},$ and $\mathtt{Q_3}$. Several edges in the join tree only have one view, which is used for all three queries.

The {\bf\color{oxfordblue}Multi-Output Optimization} layer groups the vie\-ws and output queries going out of a node such that they can be computed together over the join of the relation at the node and of its incoming views. For our running example, LMFAO groups $Q_1$, $Q_2$, and $V_{S \to I}$ because they can be computed together over the join of Sales with the incoming views $V_{T\to S}$, $V_{H\to S}$, and $V_{I\to S}$. The groups form a dependency graph as shown in Figure~\ref{fig:join-tree-favorita} (right).

LMFAO constructs a \emph{multi-output execution plan} for each group that computes all of its outgoing views and output queries in one pass over the relation at the node and using lookups into the incoming views. This is yet another instance of sharing in LMFAO: The computation of different views share the scan of the relation at the node.

The execution plan for a group is subject to fine-grained optimizations, e.g., factorized aggregate computation and shared computation. To enable them, LMFAO constructs a total order on the join attributes of the node relation. The relation and the incoming views are organized logically as tries: first grouped by the first attribute in the order, then by the next one in the context of values for the first, and so on. LMFAO then decomposes the group computation into simple arithmetic statements and lookups into incoming views that are executed at different levels in the tries.

\begin{figure}[t]
  \hspace*{-.75em}
  \begin{tikzpicture}[yscale=0.6, xscale=0.6, every node/.style={transform shape, scale=1.2}]

    \node[scale=1.3,color=black] (item) {item};
    \node[scale=1.3,color=black] at($(item)+(0,-2)$) (date) {date};
    \node[scale=1.3,color=black] at($(date)+(0,-2)$) (store) {store};

    \draw (item) -- (date) -- (store); 

    \node[scale=1.3,color=dred,anchor=west] at
    ($(item.north east)+(0.1,1ex)$) (int_item)  {$\beta_0 = 0;$};

    \node[scale=1.3,color=black,anchor=west] at
    ($(item.east)+(0.1,0)$) (int_item)  
    {foreach $i \in \pi_{\text{item}}(S\Join_{\text{item}}  V_{I \to S})$};

    \node[scale=1.3,color=dred,anchor=north west] at
    ($(int_item.south west)+(0.4,1ex)$) (agg_item1) {$\alpha_1 = V_{I\to S}(i);\hspace*{1em}{\color{oxfordblue}\alpha_2 = g(i) \cdot {\color{dred}\alpha_1};}$};
    
    \node[scale=1.3,color=dred,anchor=west] at
    ($(agg_item1.west)+(0,-.75)$) (agg_item2)  {$\beta_1 = 0;$};
    \node[scale=1.3,color=black,anchor=west] at
    ($(int_item.west)+(0.4,-2)$) (int_date) {foreach $d \in \pi_{\text{date}}(
      \sigma_{\text{item}=i}S \Join_\text{date} V_{H\to S} \Join_\text{date} V_{T\to S})$};

    \node[scale=1.3,color=dred,anchor=north west] at
    ($(int_date.south west)+(0.4,1ex)$) (agg_date1) {$\alpha_3 = V_{H\to S}(d);\hspace*{1em}{\color{oxfordblue}\alpha_4 = h(d) \cdot\alpha_2\cdot {\color{dred}\alpha_3};}$};


    \node[scale=1.3,color=dred,anchor=west] at
    ($(agg_date1.west)+(0,-.75)$) (agg_date2) {$\beta_2 = 0;$};
    
    \node[scale=1.3,color=black,anchor=west] at ($(int_date.west)+(0.4,-2)$)
    (int_store) {foreach $s \in \pi_{\text{store}}(
      \sigma_{\text{item}=i,\text{date}=d}S\Join_\text{store}\sigma_{\text{date}=d}V_{T\to S})$};

    \node[scale=1.3,color=dred,anchor=north west] at
    ($(int_store.south west)+(0.4,1ex)$) (agg_store1) {$\alpha_5 = V_{T\to S}(d,s);$};
    
    \node[scale=1.3,color=dred,anchor=west] at
    ($(agg_store1.west)+(0,-.75)$) (agg_store2) {$\beta_3 = 0; \hspace*{1em}
      {\color{oxfordblue}\alpha_6 = |\sigma_{\text{item}=i,\text{date}=d,\text{store}=s}S| \cdot \alpha_4 \cdot {\color{dred} \alpha_5};}$};
    
    \node[scale=1.3,color=black,anchor=west] at
    ($(int_store.west)+(0.4,-2)$) (int_promo)
    {$\text{foreach }u \in \pi_\text{units} \sigma_{\text{item}=i,\text{date}=d,\text{store}=s}S$};

    \node[anchor=west] at ($(int_store.west)+(0.4,-2)$) (anchor_point) {{ }};

    \node[scale=1.3,color=dred,anchor=west] at ($(anchor_point.west)+(.4,-.75)$)
    (beta_3) { $\beta_3 \pluseq u;$};
    
    \node[anchor=west] at ($(anchor_point.west)+(1,-1)$) (south_anchor) {{ }};
    
    \node[scale=1.3,color=dred,anchor=west] at ($(south_anchor.west)+(-1,-.5)$) (rs_promo)
    {$\beta_2 \pluseq \beta_3 \cdot \alpha_5;$};


    \node[scale=1.3,color=oxfordblue,anchor=west] at ($(south_anchor.west)+(-0.9,-1.1)$) (q2_out)
    {${\color{oxfordblue}\text{if } Q_2(s) \text{ then } Q_2(s) \pluseq \alpha_{6}  \text{ else } Q_2(s) = \alpha_{6};}$};

    \node[scale=1.3,color=dred,anchor=west] at ($(south_anchor.west)+(-1.4,-1.75)$)(rs_store)
    {$\beta_1 \pluseq \beta_2 \cdot \alpha_3;$};

    \node[scale=1.3,color=dred,anchor=west] at ($(south_anchor.west)+(-1.8,-2.5)$) (rs_date)
    {$\beta_0 \pluseq \beta_1 \cdot \alpha_1;$\hspace{1em} {\color{dgreen}$V_{S \to I}(i) = {\color{dred}\beta_1};$}};
    
    \node[scale=1.3,color=dred,anchor=west] at ($(south_anchor.west)+(-2.2,-3)$) (rs_item)
    {$Q_1 = \beta_0;$};
    
    \draw[color=black] ($(int_item.south west)+(0.2,0.1)$)--($(rs_item.north west)+(0.2,-0.1)$);
    \draw[color=black] ($(int_date.south west)+(0.2,0.1)$)--($(rs_date.north west)+(0.2,-0.1)$);
    \draw[color=black] ($(int_store.south west)+(0.2,0.1)$)--($(rs_store.north west)+(0.2,-0.1)$);
    \draw[color=black] ($(int_promo.south west)+(0.2,0.1)$)--($(rs_promo.north west)+(0.2,-0.1)$);
  \end{tikzpicture}
    \caption{Multi-output execution plan for {\color{dred}$Q_1$},
      {\color{oxfordblue}$Q_2$} and {\color{goodgreen}$V_{S \to I}$}.}
    \label{fig:mooexample}
\end{figure}
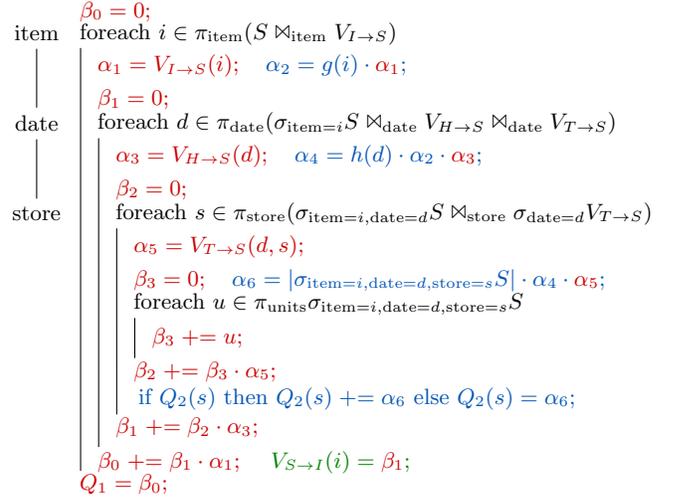

Figure~\ref{fig:mooexample} exemplifies the execution plan for Group 6 in the
dependency graph of Figure~\ref{fig:join-tree-favorita}. The attribute order for
the trie iteration is shown on the left. For simplicity of exposition, we assume
that incoming and outgoing views are functions that map tuples over their
group-by attributes to aggregates. The computation of the
outgoing views is decomposed into partial aggregates, which are pushed past loops whenever possible (loop invariant code motion) 
and stored as local variables ($\alpha$'s) or running sums ($\beta$'s). This
code optimization decreases the number of
arithmetic operations and dynamic accesses to incoming and outgoing views. 
For instance, we only look up into $V_{I\to S}$ once for each item
value and not for each (item, date, store) triple. Similarly, we only
update the result to $Q_1$ once at the very end. This optimization also allows
for sharing computation across the group. For instance, $V_{S\to I}$
shares most of its computation with $Q_1$, reflected by the
running sum $\beta_1$.

Finally, the {\bf\color{goodgreen} Code Generation} layer compiles the multi-output execution plan for each group into efficient, low-level C++ code specialized to the database schema and the join tree. 
This layer also performs low-level code optimizations, e.g., optimizing cache locality, choosing data structures for the views such as sorted arrays and (un)ordered hashmaps, and inlining function calls. LMFAO computes the groups in parallel by exploiting both task and domain parallelism.


\section{From Learning to Aggregates}
\label{sec:agg-batch}

We next show the aggregates needed for learning the three models. LMFAO computes these aggregates over the non-materialised dataset $D$, which is defined by a feature extraction query with $n$ attributes
over a multi-relational database. 

\medskip

\textbf{Linear Regression} models are linear functions:
\begin{align*}
  LR(\bm x) &= \inner{\bm \theta,\bm x} = \sum_{j \in [n]} \theta_j\cdot x_j. 
\end{align*}
with parameters $\bm \theta = ( \theta_1, \ldots,\theta_n)$ and feature vector
$\bm x = (x_1, \ldots, x_n)$.
We assume without loss of generality that (1) $x_1$ only takes
value 1 and $\theta_1$ is the intercept of the model, and (2) the label is part of the
feature vector $\bm x$ and its corresponding parameter is fixed to $-1$.

We learn the parameters $\bm\theta$ using batch gradient descent (BGD), which
requires the computation of the least-squares objective function
$J(\vec\theta)$ and its gradient $\grad J(\vec\theta)$:
\begin{align*}
  J(\vec\theta)
  &= \frac{1}{2|D|}  \bm\theta^\top \Big(\sum_{\mathbf{x}\in D} \bm x \bm x^\top\Big) \bm\theta + \frac{\lambda}{2} \norm{\vec\theta}^2\\
  \grad J(\vec\theta)
  &= \frac{1}{|D|} \Big(\sum_{\mathbf{x}\in D} \bm x \bm x^\top\Big) \bm\theta + \lambda\,\bm\theta
\end{align*}

The data-intensive computation of the optimization algorithm is given by
$\Sigma = \sum_{\mathbf{x}\in D} \bm x \bm x^\top$, which defines the
non-centered covariance matrix. The $(j,k)$-entry in $\Sigma$ accounts for the
pairwise multiplication of attributes $X_j$ and $X_k$. LMFAO computes each of
these entries as one aggregate query.

 If both $X_j$ and $X_k$ are continuous attributes,
we compute:
\begin{center}
  \texttt{SELECT SUM($\mathtt{X_j * X_k}$) FROM D}
\end{center}

Categorical attributes are one-hot encoded in a linear regression model. In
LMFAO, such attributes become group-by attributes. If only $X_j$ is
categorical, we compute:
\begin{center}
  \texttt{SELECT $\mathtt{X_j}$,SUM($\mathtt{X_k}$) FROM D GROUP BY $\mathtt{X_j}$}
\end{center}

If both $X_j$ and $X_k$ are categorical, we compute instead:
\begin{center}
  \texttt{SELECT $\mathtt{X_j,X_k}$,SUM(1) FROM D GROUP BY $\mathtt{X_j,X_k}$}
\end{center}

For the Retailer dataset, LMFAO computes 814 aggregates to learn the linear
regression model~\cite{lmfao}.  Since $\Sigma$ does not depend on the parameters
$\vec\theta$, the aggregates are computed once and then reused for all BGD
iterations.

\medskip

\textbf{Decision Trees} are popular machine learning models that
use trees with inner nodes representing conditional control statements to model
decisions and their consequences.  Leaf nodes represent predictions for the
label. We focus on learning decision trees for regression scenarios.

We learn the tree with the seminal CART algorithm~\cite{cart84}, which greedily
constructs the tree one note at a time. The algorithm learns binary trees, with
the inner nodes representing threshold conditions $X_j \texttt{ op } t$, where
$\mathtt{op}\in \{\leq, \geq, =, \neq\}$. For each node $N$, the algorithm
explores all attributes $X_j$ and possible thresholds $t_j$ to find the
condition $X_j \texttt{ op } t_j$ that minimizes the variance of the label $Y$:
\begin{align*}
  \texttt{VARIANCE}
  &= \sum_{(\mathbf{x},y) \in T} y^2 -
    \frac{1}{|T|} \Big(\sum_{(\mathbf{x},y) \in T} y\Big)^2
\end{align*}
where $T$ is the fragment of the dataset $D$ that
satisfies the condition $X_j \texttt{ op } t$ and all conditions along the path
from the root to $N$.
The algorithm thus requires the aggregates
{\tt SUM(1)}, {\tt SUM(Y)}, and {\tt SUM($\mathtt{Y^2}$)} over $T$, which can
be computed in one query over $D$:
\begin{center}
  \texttt{SELECT SUM(1),SUM(Y),SUM($\mathtt{Y^2}$) FROM D WHERE cond}
\end{center}
where \texttt{cond} is the conjunction of $X_j \texttt{ op } t$ and all
threshold conditions along to the path from root to current node.

For the Retailer dataset, LMFAO computes 3,141 aggregate queries for each node
in the decision tree~\cite{lmfao}.

\medskip

\textbf{Rk-means} computes a constant-factor
approximation of the k-means clustering objective by computing the $k$ clusters
over a small coreset of $D$~\cite{Rkmeans:AISTATS:2020}. A coreset of $D$ is a
small set of points that provide a good summarization of the original dataset
$D$. Rk-means constructs a so-called \emph{grid coreset}, which is defined as
the Cartesian product of cluster centroids computed over the projections on each
attribute of $D$.

Given the feature extraction query that defines $D$ and the constant $k$ that
defines the number of clusters, Rk-means clusters the dataset $D$ in four steps.

\emph{Step 1.}  We project $D$ onto each attribute $X_j$ and compute the weight
for each point in the projection. This can be computed as one query for each
attribute $X_j$:
\begin{center}
  \texttt{SELECT $\mathtt{X_j}$, SUM(1) FROM D GROUP BY $\mathtt{X_j}$}
\end{center}

\begin{figure*}[t]
  \begin{subfigure}[t]{0.24\textwidth}
    \includegraphics[width=1\textwidth,frame]{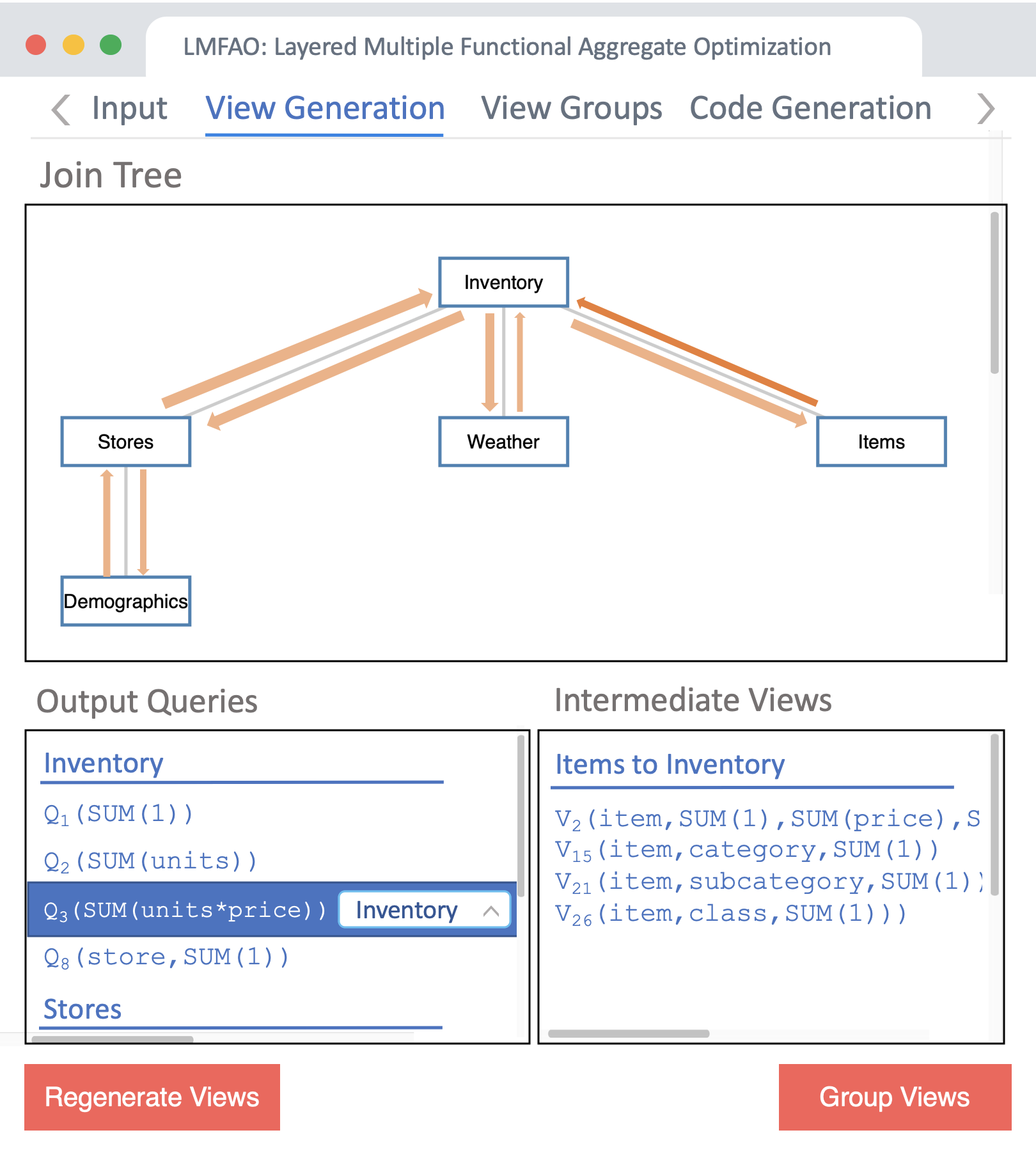}
    \caption{View Generation}
  \end{subfigure}
  \hfill 
  \begin{subfigure}[t]{0.24\textwidth}
    \includegraphics[width=1\linewidth,frame]{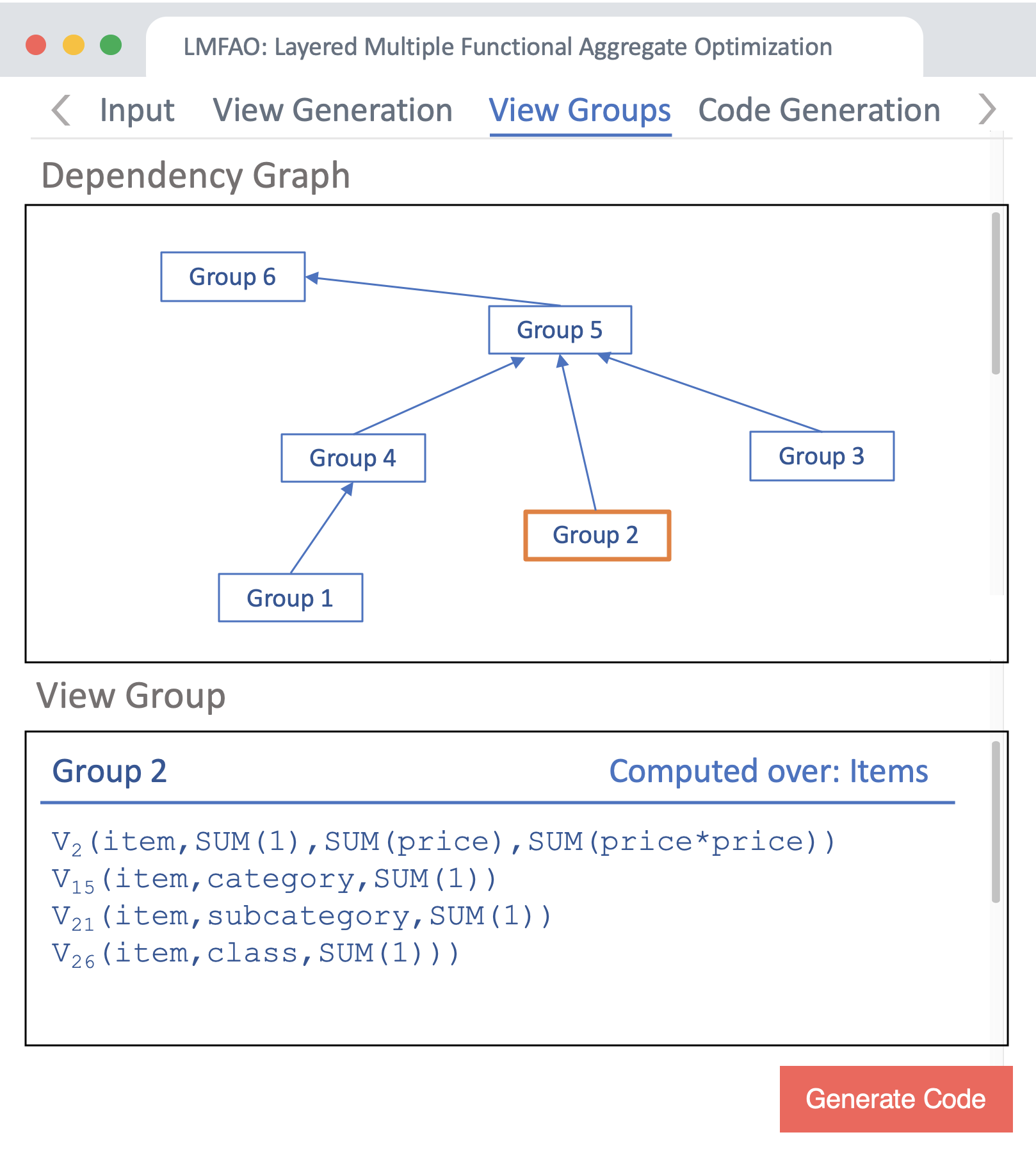}
    \caption{View Groups}
  \end{subfigure}
  \hfill 
  \begin{subfigure}[t]{0.2365\textwidth}
    \includegraphics[width=1\linewidth,frame]{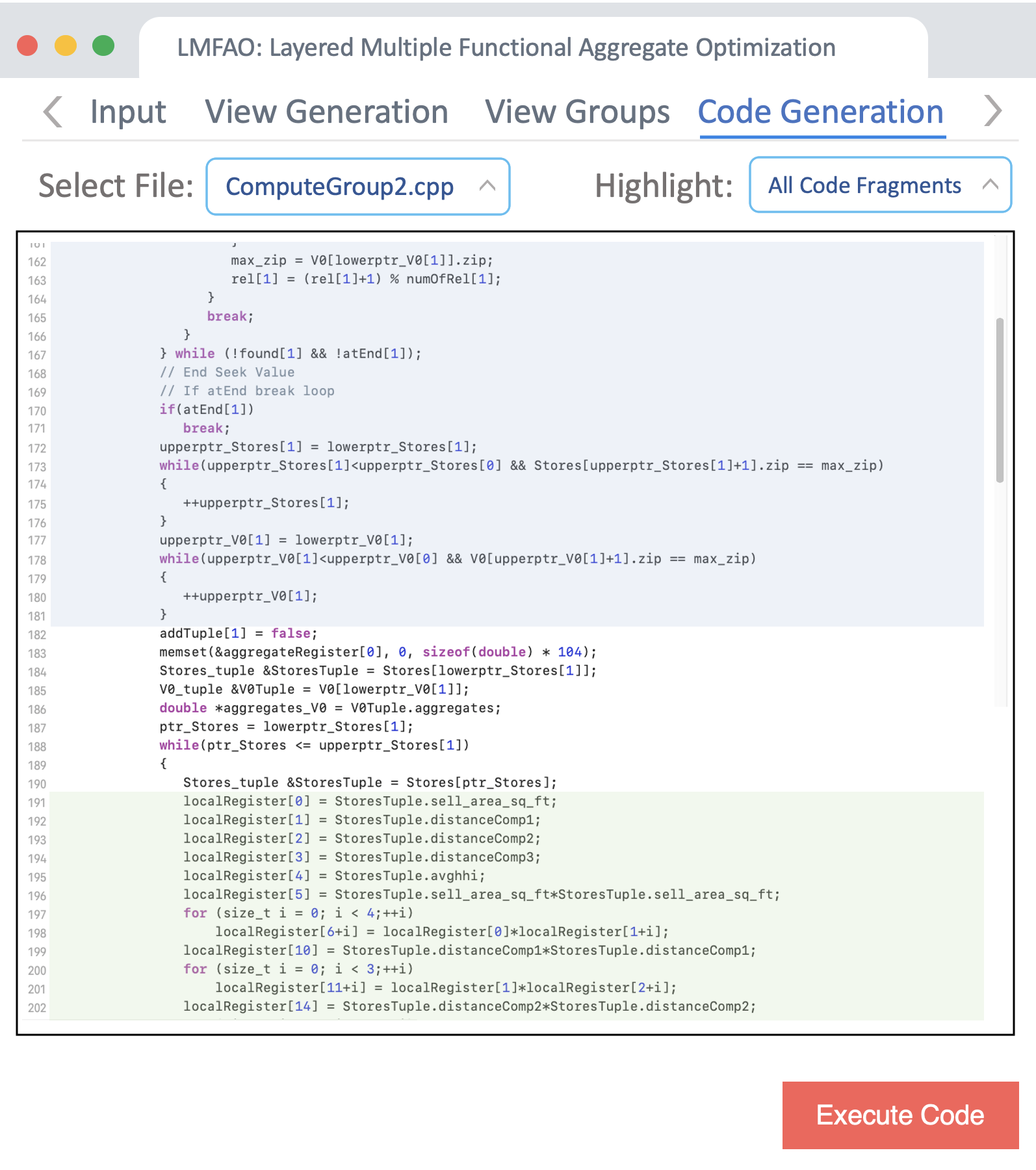}
    \caption{Code Generation}
  \end{subfigure}
  \hfill 
  \begin{subfigure}[t]{0.24\textwidth}
    \includegraphics[width=1\linewidth,frame]{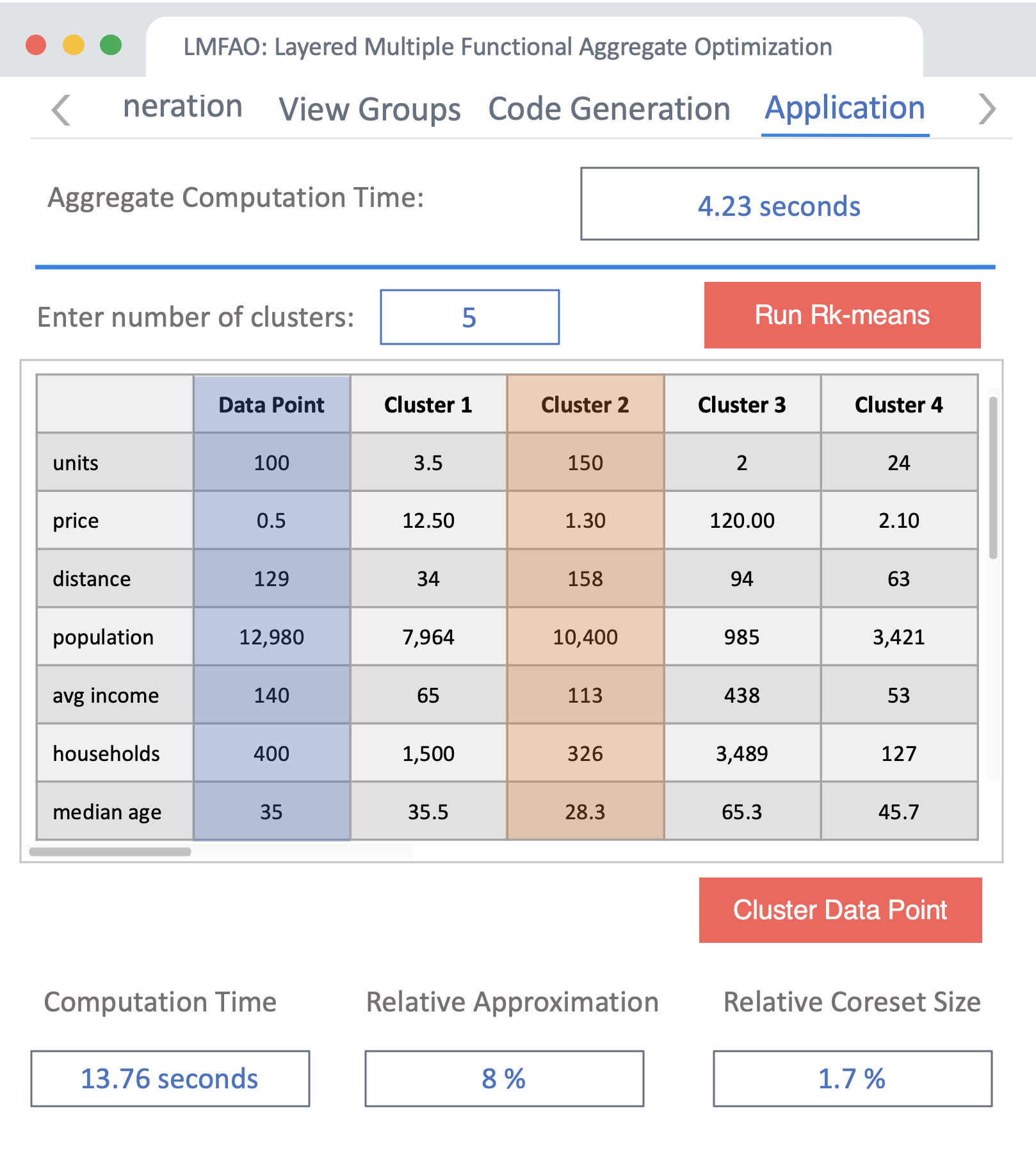}
    \caption{Rk-Means Application}
  \end{subfigure}

  \caption{Snapshots of LMFAO's user interface. Users first select a database to
    load and an ML application. Then, users can (a) inspect and modify the root
    assignment and the generated views; (b) review the grouping of views; (c)
    dive into the generated code for each view group; and (d) compute the ML
    application and analyze its performance and output.}
  \label{fig:demo}
\end{figure*}

\emph{Step 2.} We perform weighted $k$-means on each
projection.  We assume that the algorithm returns a ``cluster assignment''
relation $A_j$ which records for each $\bm x \in \pi_{X_j}(D)$ the closest
centroid $C_j$ in the projection.

\emph{Step 3.} Using the results of these clusterings we assemble a
cross-product weighted grid $G$ of centroids, which defines the coreset of
$D$. A grid point $\bm g$ in the coreset is composed of tuples of size $n$, with
the value in dimension $j\in[n]$ ranging over the possible cluster means for the
projection on $X_j$ computed in Step 2. The weight of a grid point $\bm g \in G$
is the number of data points in $D$ closest to the grid point.  The grid coreset
$G$ and the grid point weights can be computed with one aggregate query:
\begin{center}
  \texttt{SELECT $\mathtt{C_1,\ldots,C_n}$,SUM(1) FROM P GROUP BY $\mathtt{C_1,\ldots,C_n}$}
\end{center}
where $\mathtt{P = D \bowtie A_1(X_1, C_1)\bowtie \cdots A_n(X_n, C_n)}$ is the
join of $D$ and the cluster assignments from Step 2.

\emph{Step 4.} Finally, we perform weighted $k$-means clustering on the coreset
$G$ to compute the desired result of $k$ centroids.

We use LMFAO to compute steps 1 and 3 of the algorithm. This requires $n+1$
queries.


\section{Demonstration Scenarios}
\label{sec:demo}


We next describe how users can interact with LMFAO's user
interface. Figure~\ref{fig:demo} depicts snapshots of the interface.

In the \emph{Input} tab (not shown), the user chooses the databa\-se and one of
three machine learning scenarios: (1) learning linear regression models, (2)
learning regression trees, and (3) clustering using Rk-means. After
selecting the dataset, the tab depicts the join tree and database schema, so
that the user can inspect it. LMFAO then generates the batch of aggregates for
the respective application.

Next, LMFAO computes the root assignment for each aggregate query and generates
the corresponding views. The top of the \emph{View Generation} tab depicts the
join tree annotated by intermediate views, which are shown as arrows along the
edges. The width of the arrow indicates the number of views computed in this
direction. Below, the tab lists the output queries and intermediate views, where
the output queries are grouped by their root node, and intermediate views are
grouped by their directions. By default all queries and views are shown. If the
user selects a node in the join tree, only the output queries and intermediate
views that are computed over this node are listed. Similarly, by selecting one
of the arrows, only the views computed in the direction of the arrow are
shown. Figure~\ref{fig:demo} (a) depicts the selection of the arrow from Items
to Inventory.

When selecting a query in the output query list, a drop-down list for the root
of this query is shown. This allows the user to reassign the query to a
different root and change the views that are generated. The views for the new
root assignment are then regenerated and the join tree is updated. 

The \emph{View Groups} tab depicts the dependency graph of the view groups. The
user can inspect the view groups by selecting the corresponding node.

The \emph{Code Generation} tab depicts the C++ code that is generated for a
given view group. Different types of code fragments are highlighted, e.g., the
computation of the join, aggregates, or running sums. The user can choose to
highlight all code fragments, or only one of them.

The user can execute the code for the aggregate computation and inspect the
application that is computed over the aggregates. Since the execution takes a
few seconds in LMFAO, we will run it on the fly during the demonstration. Figure
\ref{fig:demo} (d) depicts the interface for Rk-means clustering, the interface
for the other two applications is similar. At the top, we show the time it took
to compute the aggregates for clustering in each dimension. The user enters the
desired number of clusters and then runs Rk-means.  Once computed, the interface
presents the cluster centroids. It also allows the user to enter the values for
a data point, and find the centroid that is closest to this point. The data
point entered in Figure~\ref{fig:demo} (d) is closest to the highlighted
centroid for Cluster 2. We further present the time it took to compute the
clusters, the relative approximation of the clusters, and the relative size of
the grid coreset with respect to the size of the dataset $D$. For the
approximation, we compute the intra-cluster distance, and take the difference
between the distances for Rk-means and the conventional Lloyd's algorithm
relative to the distance for Lloyd's. We report the average relative difference
over ten precomputed runs of Lloyd's algorithm.


\balance

\section*{Acknowledgements}

Olteanu acknowledges a research gift from Infor. Schleich is supported by a RelationalAI fellowship.  The authors acknowledge Haozhe Zhang for his help with the user interface.
This project has received funding from the European Union's Horizon 2020 research and innovation programme under grant agreement No 682588.
\bibliographystyle{abbrv}
\bibliography{bibtex}

\begin{thebibliography}{1}

\bibitem{ANNOS:DEEM:18}
M.~Abo~Khamis, H.~Ngo, X.~Nguyen, D.~Olteanu, and M.~Schleich.
\newblock {AC/DC}: In-database learning thunderstruck.
\newblock In {\em {DEEM}}, pages 8:1--8:10, 2018.

\bibitem{cart84}
L.~Breiman, J.~Friedman, R.~Olshen, and C.~Stone.
\newblock {\em {Classification and Regression Trees}}.
\newblock Wadsworth and Brooks, Monterey, CA, 1984.

\bibitem{Rkmeans:AISTATS:2020}
R.~Curtin, B.~Moseley, H.~Ngo, X.~Nguyen, D.~Olteanu, and M.~Schleich.
\newblock Rk-means: Fast clustering for relational data.
\newblock In {\em AISTATS}, pages 2742--2752, 2020.

\bibitem{favorita}
C.~Favorita.
\newblock {Corp. Favorita Grocery Sales Forecasting: Can you accurately predict
  sales for a large grocery chain?}, 2017.
\newblock
  \footnotesize\url{https://www.kaggle.com/c/favorita-grocery-sales-forecasting/}.

\bibitem{lmfao}
M.~Schleich, D.~Olteanu, M.~Abo~Khamis, H.~Ngo, and X.~Nguyen.
\newblock A layered aggregate engine for analytics workloads.
\newblock In {\em SIGMOD}, pages 1642--1659, 2019.

\bibitem{SOC:SIGMOD:16}
M.~Schleich, D.~Olteanu, and R.~Ciucanu.
\newblock Learning linear regression models over factorized joins.
\newblock In {\em SIGMOD}, pages 3--18, 2016.

\end{thebibliography}

\end{document}